\documentclass[12pt]{article}
\usepackage{axodraw,a4wide,epsfig}

\newcommand{\arccot}{\mathop{\mbox{arccot}}\nolimits}
\newcommand{\bfm}[1]{\mbox{\boldmath$#1$}}
\newcommand{\bff}[1]{\mbox{\scriptsize\boldmath${#1}$}}

\catcode`@=11
\newcount\@tempcntc
\def\@citex[#1]#2{\if@filesw\immediate\write\@auxout{\string\citation{#2}}\fi
  \@tempcnta\z@\@tempcntb\m@ne\def\@citea{}\@cite{\@for\@citeb:=#2\do
    {\@ifundefined
       {b@\@citeb}{\@citeo\@tempcntb\m@ne\@citea\def\@citea{,}{\bf ?}\@warning
       {Citation `\@citeb' on page \thepage \space undefined}}%
    {\setbox\z@\hbox{\global\@tempcntc0\csname b@\@citeb\endcsname\relax}%
     \ifnum\@tempcntc=\z@ \@citeo\@tempcntb\m@ne
       \@citea\def\@citea{,}\hbox{\csname b@\@citeb\endcsname}%
     \else
      \advance\@tempcntb\@ne
      \ifnum\@tempcntb=\@tempcntc
      \else\advance\@tempcntb\m@ne\@citeo
      \@tempcnta\@tempcntc\@tempcntb\@tempcntc\fi\fi}}\@citeo}{#1}}
\def\@citeo{\ifnum\@tempcnta>\@tempcntb\else\@citea\def\@citea{,}%
  \ifnum\@tempcnta=\@tempcntb\the\@tempcnta\else
   {\advance\@tempcnta\@ne\ifnum\@tempcnta=\@tempcntb \else \def\@citea{--}\fi
    \advance\@tempcnta\m@ne\the\@tempcnta\@citea\the\@tempcntb}\fi\fi}
\catcode`@=12
\begin{document}

\title{\vskip-3cm{\baselineskip14pt
\centerline{\normalsize DESY 02-012\hfill ISSN 0418-9833}
%\centerline{\normalsize hep-ph/0203166\hfill}
\centerline{\normalsize February 2001\hfill}}
\vskip1.5cm
Potential NRQCD and Heavy-Quarkonium Spectrum at
Next-to-Next-to-Next-to-Leading Order}
\author{\small Bernd A. Kniehl$^a$, Alexander A. Penin$^{a,b}$,
Vladimir A. Smirnov$^c$, Matthias Steinhauser$^a$\\
{\small $^a$ II. Institut f\"ur Theoretische Physik, Universit\"at Hamburg,}\\
{\small Luruper Chaussee 149, 22761 Hamburg, Germany}\\
{\small $^b$ Institute for Nuclear Research, Russian Academy of Sciences,}\\
{\small 60th October Anniversary Prospect 7a, Moscow 117312, Russia}\\
{\small $^c$ Institute for Nuclear Physics, Moscow State University,}\\
{\small 119899 Moscow, Russia}}

\date{}

\maketitle

\thispagestyle{empty}

\begin{abstract}
The next-to-next-to-next-to-leading order (N$^3$LO) Hamiltonian of potential
nonrelativistic QCD is derived.
The complete matching of the Hamiltonian and the contribution from the
ultrasoft dynamical gluons relevant for perturbative bound-state calculations
is performed including one-, two-, and three-loop contributions.
The threshold expansion is used to disentangle and match contributions of
different scales in the effective-theory calculations.
As a physical application, the heavy-quarkonium spectrum is obtained at
N$^3$LO for the case of vanishing QCD beta function.
Our results set the stage for a full N$^3$LO analysis of the heavy-quarkonium
system.

\medskip

\end{abstract}

\newpage

%%%%%%%%%%%%%%%%%%%%%%%%%%%%%%%%%%%%%%%%%%%%%%%%%%%%%%%%%%%%

\section{Introduction}
The theoretical study of nonrelativistic heavy-quark-antiquark systems
\cite{AppPol} and its applications to bottomonium \cite{NSVZ} and top-antitop
\cite{FadKho} physics rely entirely on first principles of QCD.
These systems allow for a model-independent perturbative treatment.
Nonperturbative effects \cite{Vol,Leu} are well under control for the
top-antitop system and, at least within the sum-rule approach, also for
bottomonium.
This makes heavy-quark-antiquark systems an ideal laboratory to determine
fundamental parameters of QCD, such as the strong-coupling constant $\alpha_s$
and the heavy-quark masses $m_q$.
The study of $t\bar t$ threshold production should even allow for a precision
study of Higgs-boson-induced effects.
Recently, essential progress has been made in the theoretical investigation of
the nonrelativistic heavy-quark threshold dynamics based on the
effective-theory approach \cite{CasLep,BBL}.
In such a framework, one has two expansion parameters, $\alpha_s$ and the
relative velocity $v$ of the heavy quarks.
The corrections are classified by the total power of $\alpha_s$ and $v$, i.e.\
N$^k$LO corrections contain terms of
${\cal O}\left(\alpha_s^lv^m\right)$, with $l+m=k$.
This has the consequence that, in general, different loop orders, which are
counted in powers of $\alpha_s$, contribute to the N$^k$LO result.

Analytical results for the main parameters of the nonrelativistic
heavy-quark-antiquark system are now available through the next-to-leading
order (NLO) and the next-to-next-to-leading order (NNLO)  
\cite{PinYnd,CzaMel1,BSS1,KPP,HoaTeu1,PenPiv1,PenPiv2,MelYel2}.
They have been applied to bottomonium 
\cite{PinYnd,KPP,PenPiv1,MelYel2,Hoa,BenSin}
and top-antitop \cite{HoaTeu1,PenPiv2,MelYel1,Yak,BSS2,NOS,HoaTeu2,gang}
phenomenology.
Some specific classes of the next-to-next-to-next-to-leading order (N$^3$LO)
corrections have also been investigated \cite{KniPen1,BPSV2,KniPen2,KiySum} 
(see Ref.~\cite{Pen} for a brief review).
These corrections have turned out to be so sizeable that it appears to be
indispensable to gain full control over this order, both with respect to
phenomenological applications and in order to understand the structure and
peculiarities of the nonrelativistic effective theory.
Besides its phenomenological importance, the heavy-quarkonium system is very
interesting from the theoretical point of view because it possesses a highly
sophisticated multiscale dynamics and its study demands the full power of the
effective-theory approach.
No qualitatively new theoretical effects are expected beyond N$^3$LO, so that
the N$^3$LO analysis would bring us much closer to the full understanding of
the perturbative nonrelativistic dynamics.

In the present paper, we set the stage for the complete N$^3$LO analysis of
perturbative heavy quarkonium.
In particular, we elaborate in detail the nonrelativistic effective
Hamiltonian, which is the key object of the heavy-quarkonium theory, and its
matching to the contribution associated with the emission and absorption of
dynamical ultrasoft gluons, which are relevant for perturbative bound-state
calculations \cite{KniPen1,BPSV3}.
To this end, we employ the technique of Ref.~\cite{KPSS} based on the
effective-theory approach \cite{CasLep,BBL,PinSot1} implemented with the
threshold expansion \cite{BenSmi}.

This paper is organized as follows.
In Section~\ref{sec:two}, we introduce the basic ingredients of the
nonrelativistic effective-theory formalism and describe the main features of
our approach.
In Section~\ref{sec:three}, we recall lower-order results and analyze the one-
and two-loop contributions to the N$^3$LO Hamiltonian.
In Section~\ref{sec:four}, we discuss the matching of the Hamiltonian to the
ultrasoft contribution, which necessitates the inclusion of one-, two-, and
three-loop contributions.
In Section~\ref{sec:five}, we convert our results into corrections to the
heavy-quarkonium spectrum and illustrate their phenomenological relevance.
Section~\ref{sec:six} contains our conclusions.
In the Appendix, we present some details of our calculation of the
$1/m_q$ corrections to the heavy-quark potential.

%%%%%%%%%%%%%%%%%%%%%%%%%%%%%%%%%%%%%%%%%%%%%%%%%%%%%%%%%%%%

\section{\label{sec:two}Effective theory of nonrelativistic heavy quarks}

The nonrelativistic behavior of the heavy-quark-antiquark pair is governed by
a complicated multiscale dynamics.
In the nonrelativistic regime, where the heavy-quark velocity $v$ is of the
order of the strong-coupling constant $\alpha_s$, the Coulomb effects are
crucial and have to be taken into account to all orders in $\alpha_s$.
This makes the use of the effective theory mandatory.
The effective-theory approach allows us to separate the scales and to
implement the expansion in $v$ at the level of the Lagrangian.
Let us recall that the dynamics of a nonrelativistic quark-antiquark pair is
characterized by four different {\it regions} and the corresponding modes
\cite{BenSmi}:
\begin{itemize}
\item[(i)]
the hard region (the energy and three-momentum scale like $m_q$);
\item[(ii)]
the soft region (the energy and three-momentum scale like $m_qv$);
\item[(iii)]
the potential region (the energy scales like $m_qv^2$, while the
three-momentum scales like $m_qv$); and
\item[(iv)]
the ultrasoft region (the energy and three-momentum scale like $m_qv^2$).
\end{itemize}
The ultrasoft region is only relevant for gluons, ghosts, and light quarks.
Nonrelativistic QCD (NRQCD) \cite{CasLep,BBL} is obtained by integrating out
the hard modes.
Subsequently integrating out the soft modes and the potential gluons results
in the effective theory of potential NRQCD (pNRQCD) \cite{PinSot1}, which
contains potential heavy quarks and ultrasoft gluons, ghosts, and light quarks
as active particles.
The effect of the modes that have been integrated out is two-fold:
higher-dimensional operators appear in the effective Hamiltonian,
corresponding to an expansion in $v$, and the Wilson coefficients of the
operators in the effective Hamiltonian acquire  corrections, which are series
in $\alpha_s$.

The theory of pNRQCD is relevant for the description of the heavy-quarkonium
system.
Let us recall its basic ingredients.
In pNRQCD, the (self)interactions between ultrasoft particles are described by
the standard QCD Lagrangian.
The interactions of the ultrasoft gluons with the heavy-quark-antiquark pair
are ordered in $v$ by the multipole expansion.
For the N$^3$LO analysis, only the leading-order (LO) emission and absorption
of ultrasoft gluons have to be considered.
They are described by the chromoelectric dipole interaction, which is of the
form $g_s{\bfm r}\cdot{\bfm E}$, where $g_s$ is the QCD gauge coupling,
${\bfm r}$ is the difference of coordinates of the quark and antiquark, and
${\bfm E}={\bfm E}^at^a$ is the chromoelectric field, with $t^a$ being the
generators of the colour gauge group SU(3).
The propagation of the quark-antiquark pair in the colour-singlet (s) and
colour-octet (o) states is described by the nonrelativistic Green function
$G^{s,o}$ of the Schr{\"o}dinger equation,
\begin{equation}
\left({\cal H}^{s,o}-E\right)G^{s,o}(\mbox{\boldmath $r$},
\mbox{\boldmath $r^\prime$},E)
=\delta(\mbox{\boldmath $r$}-\mbox{\boldmath $r^\prime$}),
\label{Sch}
\end{equation}
where $E$ is the energy of the quark-antiquark pair counted from the threshold
$2m_q$ and ${\cal H}^{s,o}$ is the effective nonrelativistic Hamiltonian,
\begin{eqnarray}
{\cal H}^{s,o}&=&{\cal H}^{s,o}_C+\cdots,
\nonumber\\
{\cal H}^{s,o}_C&=&-{\Delta_{\bff r}\over m_q}+V^{s,o}_C(r),
\end{eqnarray}
with $\Delta_{\bff r}=\partial_{\bff r}^2$ and $r=|{\bfm r}|$.
The ellipses stand for higher-order terms in $\alpha_s$ and $v$.
The Coulomb (C) potentials for the singlet and octet states are attractive and
repulsive, respectively, and are given by
\begin{eqnarray}
V_C^s(r)&=&-C_F\frac{\alpha_s}{r},
\nonumber\\
V_C^o(r)&=&\left(\frac{C_A}{2}-C_F\right)\frac{\alpha_s}{r},
\end{eqnarray}
where $C_A=3$ and $C_F=4/3$ are the eigenvalues of the quadratic Casimir
operators of the adjoint and fundamental representations of the colour gauge
group, respectively.
Throughout this paper, we assume that $\alpha_s=\alpha_s(\mu)$ if no argument
is specified.

The LO approximation for the Green function is given by the Coulomb solution,
which sums up terms singular at threshold and describes the leading binding
effects.
The corrections to the Coulomb Green function due to higher-order terms in the
effective  Hamiltonian can be found in Rayleigh-Schr\"odinger time-independent
perturbation theory as in standard quantum mechanics.
The Green functions have the following spectral representations:
\begin{eqnarray}
G^s({\bfm r},{\bfm r^\prime},E)
&=&\sum_{n=1}^\infty{\psi^{s*}_n({\bfm r})\psi^s_n({\bfm r^\prime})
\over E_n-E}
+\int{{\rm d}^3{\bfm k}\over(2\pi)^3}\,
{\psi^{s*}_{\bff k}({\bfm r})\psi^s_{\bff k}({\bfm r^\prime})\over
{\bfm k}^2/m_q-E},
\label{singspec}\\
G^o({\bfm r},{\bfm r^\prime},E)
&=&\int{{\rm d}^3{\bfm k}\over(2\pi)^3}\,
{\psi^{o*}_{\bff k}({\bfm r})\psi^o_{\bff k}({\bfm r^\prime})\over
{\bfm k}^2/m_q-E},
\label{octspec}
\end{eqnarray}
where $\psi_n^s$ and $\psi_{\bff k}^{s,o}$ are the wave functions of the
quark-antiquark bound and continuum states, with principal quantum number $n$
and relative three-momentum $\bfm k$, respectively, and the $E+i\varepsilon$
rule is implied.
In Eqs.~(\ref{singspec}) and (\ref{octspec}), the orbital and spin quantum
numbers, $l$ and $m$, respectively, are suppressed.
Note that a discrete part of the spectrum (bound states) only exists for the
singlet Green function.
Through the emission or absorption of an ultrasoft gluon, the quark-antiquark
pair changes its colour state, so that one switches from Eq.~(\ref{singspec})
to Eq.~(\ref{octspec}) and vice versa.

Let us now turn to the problem of perturbative calculations in the effective
theory.
Both NRQCD and pNRQCD have specific Feynman rules, which can be used for a
systematic perturbative expansion.
However, this is complicated because the expansion of the Lagrangian
corresponds to a particular subspace of the total phase space.
Thus, in a perturbative calculation within the effective theory, one has to
formally impose some restrictions on the allowed values of the virtual
momenta (see, e.g., Refs.~\cite{NioKin,AFS} for examples of highly
sophisticated calculations performed in this scheme).

Explicitly separating the phase space introduces additional scales to the
problem, such as momentum cutoffs, and makes the approach considerably less
transparent.
A much more efficient and elegant method is based on the expansion by regions
\cite{BenSmi,Smi00}, which is a systematic method to expand Feynman diagrams
in any limit of momenta and masses.
It consists of the following steps:
\begin{itemize}
\item[(i)] 
consider various regions of a loop four-momentum $k$ and expand, in every
region, the integrand in Taylor series with respect to the parameters that
are considered to be small there;
\item[(ii)]
integrate the expanded integrand over the whole integration domain of the loop
momenta; and
\item[(iii)] put to zero any scaleless integral.
\end{itemize}
In step~(ii), dimensional regularization, with $d=4-2\epsilon$ space-time
dimensions, is used to handle the divergences.
In the case of the threshold expansion in $v$, one has to deal with the four
regions and their scaling rules listed above.

In principle, the threshold expansion has to be applied to the Feynman
diagrams of full QCD.
However, after integrating out the hard modes, which corresponds to
calculating the hard-region contributions in the threshold expansion, it is
possible to apply step (i) to the diagrams constructed from the NRQCD and
pNRQCD Feynman rules \cite{KPSS}.
Equivalently, the Lagrangian of the effective theory can be employed for a
perturbative calculation without explicit restrictions on the virtual momenta
if dimensional regularization is used and the formal expressions derived from
the Feynman rules of the effective theory are understood in the sense of the
threshold expansion.
In this way, one arrives at a formulation of effective theory with two crucial
virtues: the absence of additional regulator scales and the automatic matching
of the contributions from different scales.
The second property implies that the contributions of different modes, as
computed in the effective theory, can be simply added up to get the full
result.
This automatic-matching property of effective-theory calculations in
dimensional regularization was observed in Ref.~\cite{PinSot2} and used for
high-order calculations in the theory of QED bound states in Ref.~\cite{CMY}.
We should emphasize, however, the crucial r\^ole of the threshold expansion in
effective-theory perturbative calculations because, in general, the na\"\i ve
use of the effective-theory Feynman rules and dimensional regularization leads
to incorrect results.
Another advantage of the effective-theory realization of the threshold
expansion is that the individual contributions from the hard, soft/potential,
and ultrasoft regions are manifestly gauge invariant.
Indeed, the Lagrangians of NRQCD and pNRQCD are gauge invariant, and
dimensional regularization preserves the gauge symmetry as well.
Thus, the QCD calculation of the hard corrections and the NRQCD calculation of
the soft and potential corrections to the on-shell amplitudes can be performed
in the covariant gauge suitable for relativistic problems, while the pNRQCD
calculations can be done in the Coulomb gauge appropriate for nonrelativistic
problems.
In the next sections, we illustrate the power of the approach outlined above
by an explicit analysis in pNRQCD at N$^3$LO.

%%%%%%%%%%%%%%%%%%%%%%%%%%%%%%%%%%%%%%%%%%%%%%%%%%%%%%%%%%%%

\section{\label{sec:three}Nonrelativistic effective Hamiltonian}

Let us start this section with a general remark on the structure of
higher-order corrections in pNRQCD.
As already mentioned in the Introduction, the corrections are classified by
the total power of $\alpha_s$, counting the number of loops, and $v$ or,
equivalently, $1/m_q$.
In particular, the N$^3$LO Hamiltonian includes one-loop corrections of
${\cal O}(\alpha_sv^2)$, two-loop corrections of
${\cal O}\left(\alpha_s^2v\right)$, and three-loop corrections of
${\cal O}\left(\alpha_s^3\right)$.
In NLO, the only source of corrections is the renormalization and running of
the Coulomb potential.
In NNLO, relativistic corrections in $v$ due to higher-dimensional operators
start to contribute.
In N$^3$LO, retardation effects, which cannot be described by operators of
instantaneous interaction, enter the game.
They will be discussed in the next section.
At this order, the nonrelativistic effective Hamiltonian becomes infrared (IR)
sensitive.
No qualitatively new theoretical effects are expected beyond N$^3$LO.

The general form of the Hamiltonian valid up to N$^3$LO reads
\begin{eqnarray}
{\cal H}&=&(2\pi)^3\delta({\bfm q})
\left({{\bfm p}^2\over m_q}-{{\bfm p}^4\over 4m_q^3}\right)+
C_c(\alpha_s)V_C(|{\bfm q}|)+C_{1/m}(\alpha_s)V_{1/m}(|{\bfm q}|)
+{\pi C_F\alpha_s\over m_q^2}
\nonumber\\
&&{}\times\left[C_\delta(\alpha_s)
+C_p(\alpha_s){{\bfm p}^2+{\bfm p^\prime}^2\over 2{\bfm q}^2}
+C_s(\alpha_s){\bfm S}^2+C_\lambda(\alpha_s)\Lambda({\bfm p},{\bfm q})
+C_t(\alpha_s){ T}({\bfm q})\right],
\label{Hamnnl}
\end{eqnarray}
where the following operators are involved
\begin{eqnarray}
V_C(|{\bfm q}|)&=&-\frac{4\pi C_F\alpha_s}{{\bfm q}^2},\qquad 
V_{1/m}(|{\bfm q}|)=\frac{2\pi^2 C_F\alpha_s}{m_q|{\bfm q}|},\qquad 
{\bfm S}={{\bfm \sigma}_1+{\bfm \sigma}_2\over 2},
\nonumber\\
\Lambda({\bfm p},{\bfm q})&=&
i{{\bfm S}\cdot({\bfm p}\times {\bfm q})\over {\bfm q}^2},\qquad
{ T}({\bfm q})={\bfm \sigma}_1\cdot{\bfm \sigma}_2
-3{({\bfm q}\cdot{\bfm \sigma}_1)({\bfm q}\cdot{\bfm \sigma}_2)\over
{\bfm q}^2}.
\end{eqnarray}
Here, ${\bfm p}$ and ${\bfm p^\prime}$ are the three-momenta of the incoming
and outgoing quarks, respectively, ${\bfm q}={\bfm p^\prime}-{\bfm p}$ is the
three-momentum transfer, and ${\bfm \sigma}_{1,2}$ are the quark and antiquark
spin operators.
Note that the effective Hamiltonian is defined for on-shell quarks, with
${\bfm p}^2={\bfm p^\prime}^2=m_qE$.
The Wilson coefficients are power series in $\alpha_s$,
\begin{equation}
C_i(\alpha_s)=\sum_{n=0}^\infty\left({\alpha_s\over\pi}\right)^n
c^i_n(m_q,|{\bfm q}|,\mu),
\end{equation}
where the modified minimal-subtraction ($\overline{\rm MS}$) scheme for the
renormalization of $\alpha_s$ is implied.

In the following, we discuss the nontrivial terms in Eq.~(\ref{Hamnnl}), which
are sorted in terms of the inverse heavy-quark mass. 
The contribution from the hard-virtual-momentum region is analytic in $v^2$
and starts at ${\cal O}(v^2)$.
Thus, it does not affect $V_C(|{\bfm q}|)$ and $V_{1/m}(|{\bfm q}|)$.
Corrections to the static Coulomb potential only arise from the soft
contribution.
Using renormalization group (RG) arguments, they can be rewritten as
\begin{eqnarray}
C_c(\alpha_s)V_C(|{\bfm q}|)&=&-{4\pi C_F\alpha_s(|{\bfm q}|)\over{\bfm q}^2}
\left[1+{\alpha_s(|{\bfm q}|)\over 4\pi}a_1
+\left({\alpha_s(|{\bfm q}|)\over 4\pi}\right)^2a_2\right.
\nonumber\\
&&{}+\left.\left({\alpha_s(|{\bfm q}|)\over 4\pi}\right)^3
\left(a_3+{8\pi^2}C_A^3\ln{\mu^2\over{\bfm q}^2}\right)+\cdots\right]. 
\label{Coulcor}
\end{eqnarray}
The RG logarithms can be recovered from Eq.~(\ref{Coulcor}) by recalling that
(see, e.g., Ref.~\cite{Kni})
\begin{eqnarray}
\frac{\alpha_s(|{\bfm q}|)}{\pi}&=&
\frac{\alpha_s(\mu)}{\pi}\left[1+\frac{\alpha_s(\mu)}{\pi}\beta_0L
+\left(\frac{\alpha_s(\mu)}{\pi}\right)^2L\left(\beta_0^2L+\beta_1\right)
\right.
\nonumber\\
&&{}+\left.\left(\frac{\alpha_s(\mu)}{\pi}\right)^3L
\left(\beta_0^3L^2+\frac{5}{2}\beta_0\beta_1L+\beta_2\right)+\cdots\right],
\end{eqnarray}
where $L=\ln(\mu^2/{\bfm q}^2)$ and \cite{beta}
\begin{eqnarray}
\beta_0&=&{1\over4}\left({11\over3}C_A-{4\over3}T_Fn_l\right),
\nonumber\\
\beta_1&=&{1\over16}\left({34\over3}C_A^2-{20\over3}C_AT_Fn_l-4C_FT_Fn_l
\right),
\nonumber\\
\beta_2&=&{1\over64}\left({2857\over54}C_A^3-{1415\over27}C_A^2T_Fn_l
-{205\over9}C_AC_FT_Fn_l+2C_F^2T_Fn_l+{158\over27}C_AT_F^2n_l^2
\right.\nonumber\\
&&{}+\left.{44\over9}C_FT_F^2n_l^2\right)
\end{eqnarray}
are the first three coefficients of the QCD beta function.
Here, $T_F=1/2$ is the index of the fundamental representation, and $n_l$ is
the number of light-quark flavors.
The non-RG logarithmic term of ${\cal O}({\alpha_s^3})$ in Eq.~(\ref{Coulcor})
reflects the IR divergence of the static potential \cite{ADM}.
The corresponding pole is canceled against the ultraviolet (UV) one of the
ultrasoft contribution \cite{KniPen1,BPSV1}.
For convenience, this pole is subtracted in Eq.~(\ref{Coulcor}) according to
the $\overline{\rm MS}$ prescription, so that the coefficient $a_3$ is defined
in the $\overline{\rm MS}$ subtraction scheme both for UV and IR divergences.
For consistency, the UV pole of the ultrasoft contribution has to be
subtracted in the same way.
Throughout the calculation, we use the same procedure to render the
contributions from the various regions finite.
The actual cancellation of the spurious divergences appearing in the process
of expanding by regions is reflected in the $\mu$ independence of the final
result.

In the literature \cite{KniPen1,ADM,BPSV1}, the coefficient in front of the IR 
logarithm in the ${\cal O}(\alpha_s^3)$ static potential is given as
$C^3_A/(24\pi)$, which differs from $C^3_A/(8\pi)$ in Eq.~(\ref{Coulcor}).
This is a consequence of the consistent use of dimensional regularization in
our analysis based on the threshold expansion.
The difference is due to the fact that we perform all three loop integrals in
$d$ dimensions, not just the one that is IR divergent.
The logarithmic terms not associated with IR-divergent integrals are
unphysical and are exactly canceled by similar terms from the three-loop
ultrasoft-potential-potential contribution, in which only the ultrasoft
integral is UV divergent (see Section~\ref{sec:four}), while the physical
logarithmic integral between soft and ultrasoft scales results in
$\ln\alpha_s$ corrections to the spectrum.
The calculation of the coefficients $a_i$ can be performed in the static limit
$m_q\to\infty$ of NRQCD.
Due to the exponentiation of the static potential \cite{Fis}, these
coefficients only receive contributions form the maximum non-Abelian parts.
In the language of the threshold expansion, the selection of these parts
effectively separates the contribution of the soft region.
The Abelian colour factor $C_F$ indicates the presence of the Coulomb
singularity and implies that at least one loop momentum is potential.
All such contributions are just iterations of the lower-order potential and
are taken into account in the perturbative solution of the Schr\"odinger
equation~(\ref{Sch}) around the Coulomb approximation.

The one-loop coefficient,
\begin{equation}
a_1={31\over 9}C_A-{20\over 9}T_Fn_l,
\end{equation}
has been known for a long time \cite{Fis,Bil}, while the two-loop coefficient
has only recently been found \cite{Pet,Sch}.
In our previous communication \cite{KPSS}, we confirmed the result of
Ref.~\cite{Sch},
\begin{eqnarray}
a_2&=&\left[{4343\over162}+4\pi^2-{\pi^4\over4}+{22\over3}\zeta(3)\right]C_A^2
-\left[{1798\over81}+{56\over3}\zeta(3)\right]C_AT_Fn_l
\nonumber\\
&&{}-\left[{55\over3}-16\zeta(3)\right]C_FT_Fn_l
+\left({20\over9}T_Fn_l\right)^2,
\end{eqnarray}
where $\zeta$ is Riemann's zeta function, with value
$\zeta(3)=1.202057\ldots$.
At present, only Pad\'e estimates of the tree-loop $\overline{\rm MS}$
coefficient are available, namely \cite{ChiEli} 
\begin{eqnarray}
{a_3\over 4^3}&=&
\cases{
\displaystyle
142&if $n_l=3$,\cr
\displaystyle
98&if $n_l=4$,\cr
\displaystyle
60&if $n_l=5$.\cr}
\label{a3}
\end{eqnarray}
Although the Pad\'e estimates are in reasonable agreement with the exact
results where the latter are available, the reliability of Eq.~(\ref{a3}) is
not guaranteed, and it is very desirable to exactly evaluate the coefficient
$a_3$.
However, in Section~\ref{sec:five}, we show that even a $100\%$ uncertainty in
$a_3$ would not result in a significant error in the N$^3$LO corrections to
the spectrum for the states with small principal quantum number.

The $1/m_q$-suppressed terms of Eq.~(\ref{Hamnnl}) receive contributions from 
the soft and potential regions and, applying RG techniques, can be written in
the following form:
\begin{equation}
C_{1/m}(\alpha_s)V_{1/m}(|{\bfm q}|)={\pi^2C_F\alpha_s^2(|{\bfm q}|)
\over m_q|{\bfm q}|} \left\{b_1+{\alpha_s(|{\bfm q}|)\over \pi}\left[b_2
-\frac{4}{3}\left(C_A^2+2C_AC_F\right)\ln{\mu^2\over{\bfm q}^2}\right]
+\cdots\right\}.
\label{nacor}
\end{equation}
The one-loop coefficient $b_1$ reads
\cite{GupRad,TitYnd}
\begin{equation}
b_1=-C_A+{C_F\over2}.
\label{b1}
\end{equation}
Our result for the two-loop coefficient $b_2$ is listed in Ref.~\cite{KPSS}.
A more detailed description of the calculation is presented in
Section~\ref{sec:three}.2.
The coefficient of the two-loop IR logarithm in Eq.~(\ref{nacor}) can be
extracted from the UV divergence of the ultrasoft contribution \cite{KniPen1}.
However, similarly to the case of the IR divergence of the static potential,
one has to take into account additional logarithmic terms resulting from the
consistent use of dimensional regularization.

The RG analysis results in the following representation of the $1/m_q^2$ part
of the effective Hamiltonian:
\begin{eqnarray}
\alpha_s(\mu)C_\delta(\alpha_s)&=&\alpha_s(|{\bfm q}|)
\left\{d_0^\delta+{\alpha_s(\mu)\over \pi}
\left[d_1^\delta+\frac{4}{3}(C_A-2C_F)\ln{\mu^2\over{\bfm q}^2}\right]
+\cdots\right\},
\nonumber\\
\alpha_s(\mu)C_p(\alpha_s)&=&\alpha_s(|{\bfm q}|)
\left[d_0^p+{\alpha_s(\mu)\over \pi}
\left(d_1^p-\frac{8}{3}C_A\ln{\mu^2\over{\bfm q}^2}\right)+\cdots\right],
\nonumber\\
\alpha_s(\mu)C_i(\alpha_s)&=&\alpha_s(|{\bfm q}|)
\left[d_0^i+{\alpha_s(\mu)\over \pi}d_1^i+\cdots\right],
\qquad i=s,\lambda,t,
\nonumber\\
\alpha_s(\mu)C^a_i(\alpha_s)&=&\alpha_s(m_q)
\left[d_0^{i,a}+{\alpha_s(m_q)\over\pi}d_1^{i,a}+\cdots\right],
\qquad i=\delta,s,
\label{di}
\end{eqnarray}
where the contributions from the annihilation channel are marked by the
superscript $a$.
The normalization scale of $\alpha_s$ in the one-loop scattering terms is not
fixed because they receive contributions from both the soft and hard regions.
According to the RG, the scale of $\alpha_s$ should be chosen to be
$\mu_h\approx m_q$ for the hard contribution and $\mu_s\approx\alpha_sm_q$ for
the soft one.
The IR one-loop logarithms which match the UV behavior of the ultrasoft
contribution \cite{KniPen1} are written out explicitly in Eq.~(\ref{di}).
The calculation of the one-loop coefficients $d_1^i$ within the threshold
expansion is discussed in Section~\ref{sec:three}.1.
Some of these coefficients also contain logarithms of the form
$\ln\left(m_q^2/{\bfm q}^2\right)$ originating from the logarithmic
integration between the soft and hard scales.

The purely relativistic tree-level ${\cal O}(v^2)$ corrections are given, up
to a colour factor, by the standard Breit Hamiltonian and read
\begin{eqnarray}
d_0^\delta&=&0,\qquad d_0^p=-4,\qquad d_0^s={4\over3},\qquad d_0^\lambda=6,
\qquad d_0^t={1\over3},
\nonumber\\
\qquad d_0^{\delta,a}&=&0,\qquad d_0^{s,a}=0.
\label{d0}
\end{eqnarray}

The QED effective Hamiltonian for $n_l$ light fermions is obtained from the
above expressions by setting $C_A=0$ and $C_F=T_F=1$.
Note that the QED Breit Hamiltonian has a nonvanishing one-photon annihilation
coefficient,
\begin{equation}
d_{0,\rm QED}^{s,a}=1,
\end{equation}
which is absent in the case of colour-singlet quarkonium due to colour
conservation.

In Refs.~\cite{PinSot2,GRS}, dealing with QED bound-state calculations in the
Coulomb gauge, one finds Wilson coefficients different from the Abelian parts
of those listed in Eqs.~(\ref{b1}) and (\ref{d0}).
Using them would lead to
\begin{equation}
b_1=-C_A,\qquad d_0^\delta=1.
\label{d0b1}
\end{equation}
These two sets of coefficients are equivalent, and the difference is related
to the use of off-shell operators in the Hamiltonian.
This problem is discussed in more detail in Section~\ref{sec:three}.2.

\subsection{One-loop operators}

The ${\cal O}(\alpha_sv^2)$ operators, contributing to N$^3$LO at one loop,
have attracted some attention, and an essential part of the results can be
found in  the literature
\cite{GupRad,TitYnd,GRS,BNT,PTN,Man,PinSot3,ManSte1}.
Here, we present a consistent derivation of these corrections within the
threshold-expansion framework.
The ${\cal O}(\alpha_sv^2)$ operators receive contributions from the hard and
soft/potential regions.
The contribution from the hard region requires a fully relativistic treatment.
A part of it is directly related to the Wilson coefficients $c_F$, $c_D$,
$c_S$, and $d_2$ parameterizing the Fermi, Darwin, spin-orbital, and
heavy-quark vacuum-polarization terms in the NRQCD Lagrangian, respectively
\cite{Man}.
The residual part arises from the on-shell scattering and annihilation box
diagrams at threshold \cite{PinSot3}.

The calculation of the soft contribution can be performed in NRQCD.
In the effective-theory language, we study the reduction from NRQCD to pNRQCD
and compute the effect of the soft modes being integrated out.
Apart from the standard LO terms of the NRQCD Lagrangian, we need the
$1/m_q$-suppressed terms originating from the covariant-derivative operator
${\bfm D}^2/(2m_q)$ acting on the quark and antiquark fields, the Fermi,
Darwin, and spin-orbital terms.
Note that the covariant-derivative operator includes the quark kinetic-energy
term ${\bfm k}^2/(2m_q)$.
According to the threshold expansion, it should be treated as a perturbation
if $k$ is soft or kept in the nonrelativistic quark propagator,
\begin{equation}
S(k)=\frac{1}{k_0-{\bfm k}^2/(2m_q)+i\varepsilon},
\label{prop}
\end{equation}
if $k$ is potential.
The potential region is connected with the contribution of the pole of the
propagator of Eq.~(\ref{prop}) to the integral over $k_0$.
In this connection, we can make an interesting observation.
Let us consider two-particle-irreducible diagrams that are free of Coulomb
singularities, so that the $k_0$ contour is not pinched between the poles of 
the quark and antiquark propagators.
In this case, one can close the contour of the $k_0$ integral keeping the
poles either inside or outside.
The contribution from the potential region is obviously different in these
cases, although the result for the integral is the same.
This means that separating the contributions from the soft and potential
regions for diagrams without Coulomb singularity is useless, and the
propagator of Eq.~(\ref{prop}) can thus be safely expanded in $1/m_q$ in both
regions.
In fact, the expansion of the pole contribution yields familiar generalized
functions of $k_0$, namely $\delta^{(n)}(k_0)$.
This observation dramatically simplifies the calculation, which can be
performed in a covariant form after substituting $k_0=v_0\cdot k$, where
$v_0=(1,{\bfm 0})$.

By contrast, the two-particle-reducible diagrams including the product of the
quark and antiquark propagators,
\begin{eqnarray}
\frac{1}{k_0-{\bfm k}^2/(2m_q)+i\varepsilon}\,
\frac{1}{k_0+{\bfm k}^2/(2m_q)-i\varepsilon},
\end{eqnarray}
where $k$ is the two-particle-reducible loop four-momentum, suffer from a
Coulomb singularity.
In this case, after expanding the quark propagator, one obtains ill-defined
pinched products like
\begin{eqnarray}
{1\over(k_0+i\varepsilon)^m}\,{1\over(k_0-i\varepsilon)^n}.
\end{eqnarray}
Thus, separating the soft and potential regions is unavoidable.
In the soft region, the pole contributions of the quark and antiquark
propagators have to be excluded, and the above product should be defined to be
its principal value,
\begin{eqnarray}
{1\over 2}\left[{1\over(k_0+i\varepsilon)^{m+n}}
+{1\over(k_0-i\varepsilon)^{m+n}}\right],
\end{eqnarray}
which again allows for a covariant treatment.
In the potential region, the quark and antiquark propagator poles produce
contributions of the form
\begin{eqnarray}
-i\pi\,{m_q\over{\bfm k}^2-i\varepsilon}
\left[\delta\left(k_0-{{\bfm k}^2\over2m_q}\right)
+\delta\left(k_0+{{\bfm k}^2\over2m_q}\right)\right],
\label{pole}
\end{eqnarray}
where the $1/v$ Coulomb singularity shows up explicitly.
After integration over $k_0$, Eq.~(\ref{pole}) yields the nonrelativistic
Green function of the free Schr\"odinger equation.
A contribution of this type can always be related to iterations of the
operators of the effective Hamiltonian in time-independent perturbation
theory.
Therefore, it should not be considered as a correction to the Hamiltonian.
However, there is one subtle question here, namely as to whether the
operators that vanish for on-shell quarks should be included in the effective
Hamiltonian or not.
This does not affect the $1/m_q^2$ potential, but it matters for the $1/m_q$
operator discussed in Section~\ref{sec:three}.2.

One has to be careful with the definition of commutators of the Dirac/Pauli
matrices within dimensional regularization.
Since poles in $\epsilon$ are only present in the individual contributions 
from the hard and soft regions and drop out in the sum, one can explicitly
retain the commutators during the analysis of the these regions and replace
them by the four/three-dimensional expressions in the final result \cite{CMY}.
Otherwise one has to use the same prescription for the evaluation of the
commutators of the Dirac/Pauli matrices in the hard and soft regions.
Throughout the calculation, we use the four/three-dimensional antisymmetric
$\epsilon$ tensor for the definition of the commutators as was done in
Ref.~\cite{ManSte1}.
(This differs from the prescription of Ref.~\cite{PinSot3}.)

The one-loop calculation poses no technical problems, and our results for the
Wilson coefficients read
\begin{eqnarray}
\lefteqn{d_1^\delta+\frac{4}{3}(C_A-2C_F)\ln{\mu^2\over{\bfm q}^2}
=\left[\left(-\frac{1}{4}-\frac{17}{6}\ln{\mu^2\over m_q^2}\right)C_A
+\left(\frac{5}{3}-\frac{1}{3}\ln{\mu^2\over m_q^2}\right)C_F
-{4\over15}T_F\right]_h}
\nonumber\\
&&{}+\left[\left(\frac{9}{4}+\frac{25}{6}\ln{\mu^2\over{\bfm q}^2}\right)C_A
+\left(\frac{1}{3}-\frac{7}{3}\ln{\mu^2\over{\bfm q}^2}\right)C_F\right]_s,
\nonumber\\
&&d_1^p-\frac{8}{3}C_A\ln{\mu^2\over{\bfm q}^2}
=\left[\left(-{31\over9}-\frac{8}{3}\ln{\mu^2\over{\bfm q}^2}\right)C_A
+{20\over9}T_Fn_l\right]_s,
\nonumber\\
&&d_1^s=\left[\left({4\over3}+{7\over6}\ln{\mu^2\over m_q^2}\right)C_A
-{2\over3}C_F\right]_h
+\left[\left(-{14\over27}-{7\over6}\ln{\mu^2\over{\bfm q}^2}\right)C_A
-{20\over27}T_Fn_l\right]_s,
\nonumber\\
&&d_1^\lambda=\left[\left(4+2\ln{\mu^2\over m_q^2}\right)C_A+4C_F\right]_h
+\left[\left({7\over6}-2\ln{\mu^2\over{\bfm q}^2}\right)C_A
-{10\over3}T_Fn_l\right]_s,
\nonumber\\
&&d_1^t=\left[\left({1\over3}+{1\over6}\ln{\mu^2\over m_q^2}\right)C_A
+{1\over3}C_F\right]_h
+\left[\left({13\over108}-{1\over6}\ln{\mu^2\over{\bfm q}^2}\right)C_A
-{5\over27}T_Fn_l\right]_s,
\nonumber\\
&&d_1^{\delta,a}=(-4+4\ln2-i2\pi)T_F,
\nonumber\\
&&d_1^{s,a}=(2-2\ln2+i\pi)T_F,
\label{d1}
\end{eqnarray}
where the contributions from the hard $(h)$ and soft $(s)$ regions are
explicitly separated.
The first two equations of Eq.~(\ref{d1}) are written in a way appropriate for
Eq.~(\ref{di}).
As in NNLO, the one-photon-annihilation channel provides an additional
contribution to the QED Wilson coefficient, namely
\begin{eqnarray}
d_{1,\rm QED}^{s,a}&=&2-2\ln2+i\pi
-\left({5\over3}-2\ln2+i\pi\right){n_l\over3}.
\end{eqnarray}
The QED part of Eq.~(\ref{d1}) is in agreement with Ref.~\cite{PinSot1}.
The non-Abelian part of Eq.~(\ref{d1}) agrees with Ref.~\cite{ManSte1},
including the nonlogarithmic $C_A$ term in the coefficient $d^\delta_1$,
which differs from the one of Ref.~\cite{TitYnd}.

\subsection{Two-loop operators}

The ${\cal O}\left(\alpha_s^2v\right)$ part or the effective Hamiltonian is
given by the two-loop corrections to the $1/m_q$ potential.
These corrections are solely generated by the covariant-derivative operator in
the NRQCD Hamiltonian.
The calculation of the two-loop $1/m_q$ potential is simplified by the absence
of the hard contribution, but, in turn, it is complicated by the presence of
off-shell operators. 

To introduce the problem, let us start with the one-loop $1/m_q$ potential.
For illustrative purposes, we use the Feynman gauge, where the Coulomb and
transverse gluons do not mix.
Nonzero contributions come from the planar and crossed box diagrams and a
diagram with one three-gluon vertex.
The last two diagrams do not contain Coulomb singularities.
According to the procedure described in the previous section, the quark and
antiquark propagators should be expanded in $1/m_q$.
The result reads
\begin{equation}
-\left(C_AC_F-C_F^2\right)\frac{\pi^2\alpha_s^2}{m_q|{\bfm q}|}.
\label{cros}
\end{equation}
The soft contribution to the planar box diagram vanishes because the $1/m_q$ 
terms from the expansion of the quark and antiquark propagators cancel.
The potential contribution corresponds to the second iteration of the
operators generated by the exchange of one potential gluon.
The operators that are defined for on-shell quarks enter the effective
Hamiltonian, and their iteration is taken into account when Eq.~(\ref{Sch}) is
solved.
However, the potential-gluon exchange also generates operators proportional to
the energy transfer $q_0=({\bfm p^\prime}^2-{\bfm p}^2)/(2m_q)$, which vanish
for on-shell quarks.
Such operators do not enter the effective Hamiltonian, which is defined
on-shell, and their iterations should be considered as corrections to the
effective Hamiltonian.
The factor $q_0$ cancels the denominator of the free nonrelativistic Green
function and effectively makes the diagram two-particle irreducible.
The Coulomb singularity brings a factor of $m_q$, so that for the calculation
of the $1/m_q$ corrections we need the iteration of the $1/m_q^2$ off-shell
operator and the leading Coulomb potential.
In the case under consideration, the relevant off-shell operator is
\begin{equation}
-\frac{\pi C_F\alpha_s}{m_q^2}
\left(\frac{{\bfm p^\prime}^2-{\bfm p}^2}{{\bfm q}^2}\right)^2.
\label{offshell}
\end{equation}
Explicit evaluation of the corresponding potential contribution yields
\begin{equation}
-\frac{\pi^2C_F^2\alpha_s^2}{2m_q|{\bfm q}|},
\label{onshell}
\end{equation}
and the coefficients of Eqs.~(\ref{cros}) and (\ref{onshell}) sum up to
Eq.~(\ref{b1}).
In fact, by using the Coulomb equation of motion, it is straightforward to
check that the matrix elements of Eqs.~(\ref{offshell}) and (\ref{onshell})
between Coulomb states are the same.

Note that, in QED calculations performed in the Coulomb gauge
\cite{PinSot2,GRS}, the off-shell tree-level operator analogous to
Eq.~(\ref{offshell}) naturally appears.
If one includes this operator in the effective Hamiltonian, the Wilson
coefficient $d_0^\delta$ is given by  Eq.~(\ref{d0b1}) and the coefficient
$b_1$ is purely non-Abelian.

The equivalence of the two formulations is obvious from the above analysis.
The use of off-shell operators is advantageous in QED because it allows one to
reduce the number of loops by means of the Coulomb equation of motion, as may
be seen by comparing Eqs.~(\ref{offshell}) and (\ref{onshell}).
However, here we use the on-shell formulation and the general covariant gauge,
which is more suitable for multiloop QCD calculations.

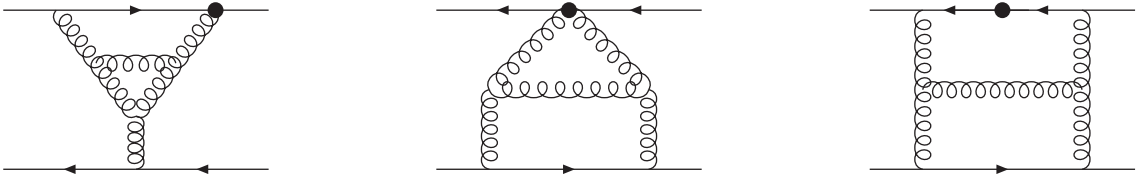
\begin{figure}[ht]
\vspace{2em}
\begin{center}
\begin{picture}(100,60)(60,0)
\ArrowLine(0,60)(100,60)
\ArrowLine(50,0)(0,0)
\ArrowLine(100,0)(50,0)
%\ArrowLine(65,40)(50,20)
%\ArrowLine(50,20)(35,40)
%\ArrowLine(35,40)(65,40)
\Gluon(50,20)(20,60){3}{9}
\Gluon(35,40)(65,40){3}{4.5}
\Gluon(50,20)(50,0){3}{4.5}
\Gluon(80,60)(50,20){3}{9}
%\Gluon(35,40)(20,60){3}{4.5}
\Vertex(80,60){3}
\end{picture}
\begin{picture}(100,60)(0,0)
\ArrowLine(0,0)(100,0)
\ArrowLine(50,60)(0,60)
\ArrowLine(100,60)(50,60)
%\GlueArc(50,30)(30,0,90){3}{6.5}
%\GlueArc(50,30)(30,90,180){3}{6.5}
\Gluon(20,30)(50,60){3}{6.5}
\Gluon(50,60)(80,30){3}{6.5}
\Gluon(20,0)(20,30){3}{5.5}
\Gluon(80,30)(80,0){3}{5.5}
\Gluon(80,30)(20,30){3}{8.5}
\Vertex(50,60){3}
\end{picture}
\begin{picture}(100,60)(-60,0)
\ArrowLine(0,0)(100,0)
\ArrowLine(60,60)(0,60)
\ArrowLine(100,60)(30,60)
\Gluon(20,30)(80,30){3}{10.5}
\Gluon(20,0)(20,60){3}{10.5}
\Gluon(80,60)(80,0){3}{10.5}
\Vertex(50,60){3}
\end{picture}
\vspace{1em}
\caption{\small Examples of two-particle-irreducible two-loop diagrams.
The standard quark-gluon vertex represents the leading Coulomb interaction.
The black circles correspond to the three types of ${\cal O}(1/m_q)$ terms
generated by the quark covariant-derivative term.}
\label{fig:one}
\end{center}
\end{figure}

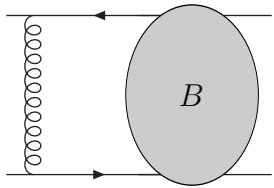
\begin{figure}[ht]
\vspace{2em}
\begin{center}
\begin{picture}(100,70)(0,0)
\ArrowLine(0,0)(70,0)
\ArrowLine(70,60)(0,60)
\ArrowLine(50,0)(100,0)
\ArrowLine(100,60)(50,60)
\GOval(70,30)(34,25)(0){0.8}
\Gluon(10,0)(10,60){3}{10.5}
\Text(70,30)[cc]{$B$}
\end{picture}
\vspace{1em}
\caption{\small Example of a two-particle-reducible two-loop diagram.
$B$ stands a general one-loop two-particle-irreducible subgraph.
The threshold expansion of this diagram is discussed in the text.}
\label{fig:two}
\end{center}
\end{figure}

The structure of the expansion remains intact at two loops, and our final
result reads
\begin{equation}
b_2=-\left(\frac{101}{36}+\frac{4}{3}\ln{2}\right)C_A^2
+\left(\frac{65}{18}-\frac{8}{3}\ln2\right)C_AC_F+\frac{49}{36}C_AT_Fn_l
-\frac{2}{9}C_FT_Fn_l.
\label{b2}
\end{equation}
There is no fully Abelian $C_F^3$ contribution in $b_2$, as is known from the
QED analysis \cite{PinSot2,GRS}.
In the calculation of the two-loop two-particle-irreducible diagrams, which
completely determine the maximum non-Abelian $C_A^2C_F$ and $C_AC_FT_Fn_l$
structures of the result, we used an expanded form of Eq.~(\ref{prop}).
Typical diagrams contributing to the $C_A^2C_F$ part are depicted in 
Fig.~\ref{fig:one}.
The analysis of the relevant two-particle-reducible diagrams, shown in
Fig.~\ref{fig:two}, is conceptually similar to the one-loop case described
above.
Let us discuss it in more detail.
The reducible loop momentum can be either soft or potential.
If it is soft, the quark and antiquark propagators can be expanded, and the
only nonzero contribution corresponds to the situation where the single gluon
is the Coulomb one and the $1/m_q$ term is kept in the expansion of the
one-loop block, $B$.
If the reducible loop momentum is potential, then one only has to take into
account contributions from the off-shell operator.
There are two possibilities:
(i) the off-shell operator is generated by the single-gluon exchange, and the
block $B$ stands for the one-loop corrections to the Coulomb potential; or
(ii) the off-shell operator comes from the $1/m_q^2$ part of the one-loop
block $B$, and the single gluon is the Coulomb one.
The analysis of the potential contribution is rather straightforward from the
technical point of view.
The calculation of the two-particle-irreducible diagrams and the soft parts of
the two-particle-reducible diagrams is more involved.
Some details are presented in the Appendix.

We performed a number of nontrivial checks for our analysis.
(i) We worked in the general covariant gauge and verified that the gauge
parameter cancels in our final result.
(ii) The two-loop expression from which Eq.~(\ref{b2}) is obtained contains
both UV and IR divergences.
The UV ones were removed in Eq.~(\ref{b2}) by the renormalization of
$\alpha_s$ in the one-loop result of Eq.~(\ref{nacor}).
On the other hand, the IR divergences were canceled by the UV ones of the
ultrasoft contribution (see Section~\ref{sec:four}) leaving a finite result
for the spectrum.
The RG logarithms proportional to $\beta_0$ and the IR logarithms are in
agreement with Eq.~(\ref{nacor}).
(iii) To test our program, we also recalculated the two-loop corrections to
the static heavy-quark-antiquark potential and found agreement with
Ref.~\cite{Sch}.
Note that, with our prescription for the calculation of the soft and potential
contributions, we explicitly obtained zero for the partially Abelian
corrections to the static potential.

%%%%%%%%%%%%%%%%%%%%%%%%%%%%%%%%%%%%%%%%%%%%%%%%%%%%%%%%%%%%

\section{\label{sec:four}Ultrasoft contribution}

\begin{figure}[ht]
\vspace{2em}
\begin{center}
\begin{picture}(300,60)(0,8)
\GBox(50,8)(100,12){0.5}
\GBox(200,8)(250,12){0.5}
\Line(100,12)(200,12)
\Line(100,8)(200,8)
\GlueArc(150,10)(50,0,180){3.5}{22.5}
\Vertex(100,10){3.5} \Vertex(200,10){3.5}
\Text(150,60)[cb]{}
\end{picture}
\vspace{1em}
\caption{\small Feynman diagram giving rise to the ultrasoft contribution at
N$^3$LO. 
The shaded and light double lines stand for the singlet and octet Green
functions, respectively.
The loopy line represents the ultrasoft-gluon propagator in the Coulomb gauge,
and the black circles correspond to the chromoelectric dipole interaction.}
\label{fig:three}
\end{center}
\end{figure}
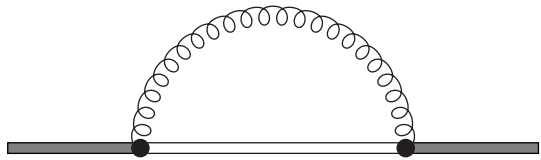

For the N$^3$LO Hamiltonian, only the leading retardation effects are needed.
They arise from the chromoelectric dipole interaction of the heavy quarkonium
with a virtual ultrasoft gluon, as depicted in Fig.~\ref{fig:three}.
In the analysis of the ultrasoft contribution, we proceed along the lines of
the original analysis \cite{KniPen1}.
As has been mentioned in Section~\ref{sec:two}, there is freedom in the
choice of gauge.
We work in the Coulomb gauge, which is especially appropriate for N$^3$LO
calculations in pNRQCD because the Coulomb gluon does not propagate and the
dynamical gluon is transverse.

The analytical expression for the corrections to the singlet Coulomb wave
function can be obtained by using the pNRQCD Feynman rules of 
Section~\ref{sec:two}.
It reads
\begin{equation}
\delta G^s({\bfm x},{\bfm y},E)=
-\sum_{~m}\hspace{-5mm}\int\sum_{~n}\hspace{-5mm}\int
{\psi^*_m({\bfm x})\psi_n({\bfm y})\over(E_m-E)(E_n-E)}J_{mn}(E),
\label{Gfcorrus}
\end{equation}
where
\begin{equation}
J_{mn}(E)=-C_Fg_s^2(\mu_{\rm us})\int{{\rm d}^3{\bfm k}\over(2\pi)^3}
\langle r_j\rangle_{m{\bff k}}\langle r_i\rangle_{{\bff k}n}
I^{ij}\left(E-{{\bfm k}^2\over m_q}\right),
\label{jmn}
\end{equation}
with
\begin{equation}
I^{ij}(t)=-i\int{{\rm d}^dl\over(2\pi)^d}\,
{l_0^2(\delta^{ij}-l^il^j/{\bfm l^2})
\over l^2(t-l^0)}.
\label{imnrep}
\end{equation}
The sum/integral in Eq.~(\ref{Gfcorrus}) goes over the whole spectrum, and
$m$ and $n$ stand for the complete set of quantum numbers characterizing the
discrete/continuum part of the spectrum.
The scale $\mu_{\rm us}\approx\alpha_s^2m_q$ of $g_s$ in Eq.~(\ref{jmn})
reflects the ultrasoft-momentum flow through the gluon propagator.
The matrix element $\langle{\bfm r}\rangle_{{\bff k}n}$ is taken between the
singlet Coulomb wave function of quantum number $n$ and the octet Coulomb wave
function of three-momentum ${\bfm k}$.
Performing in Eq.~(\ref{imnrep}) the integration over $l_0$, we recover the
expression of time-independent nonrelativistic perturbation theory.
The remaining integral over ${\bfm l}$ is UV divergent.
Subtracting the UV pole according to the $\overline{\rm MS}$ prescription, we
obtain
\begin{equation}
I^{ij}(E-{\bfm k}^2/m_q)={\delta^{ij}\over6\pi^2}
\left(E-{{\bfm k}^2\over m_q}\right)^3\left(
\ln{\left|E_1^C\right|\over{\bfm k}^2/m_q-E}
+\ln{\mu\over\left|E_1^C\right|}+{5\over6}-\ln2\right),
\label{imnres}
\end{equation}
where, for convenience, the Coulomb energy,
\begin{eqnarray}
E^C_n&=&-{C_F^2\alpha_s^2m_q\over4n^2},
\end{eqnarray}
with $n=1$, has been introduced into the arguments of the logarithms.
The $\bfm k$-dependent logarithmic term in Eq.~(\ref{imnres}) represents a
pure retardation effect and cannot be interpreted in terms of some
instantaneous interaction.
It receives contributions from Coulomb-gluon-exchange diagrams of all orders
and leads to a QCD analogue of the familiar Bethe logarithms in the
corrections to the spectrum, to be discussed in the next section.
On the other hand, making use of the completeness relation,
\begin{equation}
\int{{\rm d}^3{\bfm k}\over(2\pi)^3}
\langle r_j\rangle_{m{\bff k}}\langle r_i\rangle_{{\bff k}n}
\left(E-{{\bfm k}^2\over m_q}\right)^l
=\left\langle{\bfm r}(E-{\cal H}_C^o)^l{\bfm r}\right\rangle_{mn},
\label{comrel}
\end{equation}
the remaining part of Eq.~(\ref{imnres}), excluding the $\bfm k$-dependent
logarithmic term, is reduced to an instantaneous interaction of the form
\begin{eqnarray}
\left\langle{\bfm r}\left(E-{\cal H}_C^o\right)^3{\bfm r}\right\rangle_{mn}
&=&\left\langle-{C_A^3\over 8}\,{\alpha_s^3\over r}
-\left(C_A^2+2C_AC_F\right){\alpha_s^2\over m_qr^2}
+4(C_A-2C_F){\pi\alpha_s\over m_q^2}\delta({\bfm r})\right.
\nonumber\\
&&{}+\left.C_A\frac{\alpha_s}{m_q^2}
\left\{\Delta_{\bfm r},{1\over r}\right\}
\right\rangle_{mn}+\mbox{reducible part},
\label{matel}
\end{eqnarray}
where the reducible part includes terms with the operator $E-{\cal H}^s_C$
acting directly on a wave function.
By using Eqs.~(\ref{Gfcorrus}), (\ref{imnres}), (\ref{comrel}), and
(\ref{matel}), one arrives at the following representation of the corrections
to the Green function:
\begin{equation}
\delta G^s({\bfm r},{\bfm r^\prime},E)
=-\int{\rm d}^3{\bfm r^{\prime\prime}}
G^s_C({\bfm r},{\bfm r^{\prime\prime}},E)
{\cal H}^{\rm us}(r^{\prime\prime})
G^s_C({\bfm r^{\prime\prime}},{\bfm r^\prime},E)
+\mbox{contact terms},
\label{corrinst}
\end{equation}
where, in momentum space,
\begin{eqnarray}
{\cal H}^{\rm us}&=&{C_F\alpha_s\over3}
\left(\frac{1}{2}\ln{\mu^2\over\left(E_1^C\right)^2}+{5\over6}-\ln2\right)
\left[C_A^3{\alpha_s^3\over{\bfm q}^2}
+4\left(C_A^2+2C_AC_F\right){\pi\alpha_s^2\over m_q|{\bfm q}|}\right.
\nonumber \\
&&{}-\left.8(C_A-2C_F){\alpha_s\over m_q^2}
+16C_A{\alpha_s\over m_q^2}\,{{\bfm p}^2
+{\bfm p^\prime}^2\over2{\bfm q}^2}\right].
\label{Hamus}
\end{eqnarray}
The first term in Eq.~(\ref{corrinst}) imitates the corrections to the Green
function due to the N$^3$LO term of Eq.~(\ref{Hamus}) in the effective
Hamiltonian.
The contact terms correspond to the reducible part of Eq.~(\ref{matel}).
By the equation of motion (\ref{Sch}), one or both Coulomb Green functions in
Eq.~(\ref{corrinst}) are converted into $\delta$ functions.
The corresponding contribution cannot be imitated by a term of the effective
Hamiltonian.
It does not affect the energy levels, but it leads to corrections to the
wave functions.
A part of such corrections, namely the on-shell renormalization of the
heavy-quarkonium wave function at the origin, was computed in
Ref.~\cite{KniPen1}.
In the present paper, we refrain from discussing this type of corrections.

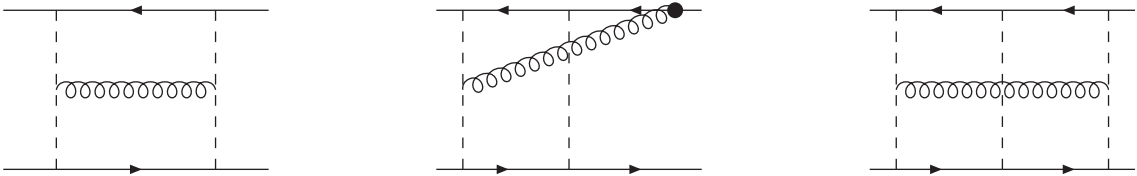
\begin{figure}[ht]
\vspace{2em}
\begin{center}
\begin{picture}(100,60)(60,0)
\ArrowLine(0,0)(100,0)
\ArrowLine(100,60)(0,60)
\DashLine(80,60)(80,0){4}
\DashLine(20,0)(20,60){4}
\Gluon(20,30)(80,30){3}{10.5}
\end{picture}
\begin{picture}(100,60)(0,0)
\ArrowLine(0,0)(50,0)
\ArrowLine(50,60)(0,60)
\ArrowLine(50,0)(100,0)
\ArrowLine(100,60)(50,60)
\Gluon(10,30)(90,60){3}{14.5}
\DashLine(10,0)(10,60){4}
\DashLine(50,0)(50,60){4}
\Vertex(90,60){3}
\end{picture}
\begin{picture}(100,60)(-60,0)
\ArrowLine(0,0)(50,0)
\ArrowLine(50,60)(0,60)
\ArrowLine(50,0)(100,0)
\ArrowLine(100,60)(50,60)
\Gluon(10,30)(90,30){3}{14.5}
\DashLine(10,0)(10,60){4}
\DashLine(50,0)(50,60){4}
\DashLine(90,0)(90,60){4}
\end{picture}
\vspace{1em}
\caption{\small Examples of two- and three-loop diagrams encoded in the
diagram of Fig.~\ref{fig:three}, which require additional matching to bring
Eq.~(\ref{Hamus}) in agreement with the threshold expansion.
The dashed and loopy lines represent the potential (Coulomb) and ultrasoft
(transverse) gluon propagators in the Coulomb gauge, respectively.
The black circles correspond to the interaction generated by the quark
covariant-derivative term.}
\label{fig:four}
\end{center}
\end{figure}

Although dimensional regularization was used in deriving Eq.~(\ref{Hamus}),
the latter is not consistent with the threshold expansion because the
three-dimensional expression for the Coulomb Green function was used instead
of the $(d-1)$-dimensional one, which is not available.
Thus, Eq.~(\ref{Hamus}) derived in Ref.~\cite{KniPen1} requires some
additional matching.
For this purpose, we separate the divergent contributions from the diagram
of Fig.~\ref{fig:three} and compute them according to the threshold expansion.
Eq.~(\ref{Hamus}) implies that only the one-, two-, and three-loop 
contributions encoded in the diagram of Fig.~\ref{fig:three} are divergent.
They include one divergent ultrasoft loop integration and zero, one, or two
convergent potential loop integrations.
The three-dimensional form of the Coulomb Green function used in the
evaluation above implies that the integration over the potential momenta is
performed in three dimensions, while, for the corresponding contributions
obtained within the threshold expansion, they are done in $d-1$ dimensions.
Thus, the matching term is given by the difference of the diagrams computed
in $d-1$ dimensions and the same diagrams with three-dimensional integrals
over the potential momenta in the limit $\epsilon\to0$.
Only two- and three-loop diagrams have to be considered.
Examples of such diagrams are presented in Fig.~\ref{fig:four}.
The calculation is simplified by the fact that, for the matching, we only need
the pole part of the ultrasoft integral, which factorizes.
For the two-particle-reducible diagrams of the type shown in
Fig.~\ref{fig:two}, one has to take into account the off-shell operators
generated by the two-particle-irreducible block $B$ with one ultrasoft loop in
a way similar to the case of the $1/m_q$ potential in
Section~\ref{sec:three}.2.
In addition to the ${\cal O}(\alpha_sv^2)$ operator proportional to
Eq.~(\ref{offshell}) with an extra factor of $\alpha_s$, the one-loop
ultrasoft exchange generates an off-shell operator proportional to
\begin{eqnarray}
\frac{\pi C_F\alpha_s}{m_q^2}\,
\frac{{\bfm p}^2+{\bfm p^\prime}^2-2m_qE}{{\bfm q}^2},
\end{eqnarray}
which, in three dimensions after adding one extra Coulomb gluon, results in
the same $1/m_q$ term as the operator of Eq.~(\ref{offshell}).
The matching terms are found to be
\begin{eqnarray}
\delta{\cal H}^{\rm us}&=&{C_F\alpha_s\over3}
\left\{\ln{\mu^2\over{\bfm q}^2}C_A^3{\alpha_s^3\over{\bfm q}^2}
+\left[\left(1+4\ln2+2\ln{\mu^2\over{\bfm q}^2}\right)C_A^2
\right.\right.
\nonumber\\
&&{}-\left.\left.
4\left(2-2\ln2-\ln{\mu^2\over{\bfm q}^2}\right)C_AC_F\right]
{\pi\alpha_s^2\over m_q|{\bfm q}|}\right\}.
\end{eqnarray}
Incidentally, the three-loop coefficient does not have a constant term.
The $\mu$ dependence of the ultrasoft contribution
${\cal H}^{\rm us}+\delta{\cal H}^{\rm us}$ exactly cancels the $\mu$
dependence of the N$^3$LO Hamiltonian given in Eqs.~(\ref{Coulcor}),
(\ref{nacor}), and (\ref{di}) ensuring the cancellation of IR and UV poles.

%%%%%%%%%%%%%%%%%%%%%%%%%%%%%%%%%%%%%%%%%%%%%%%%%%%%%%%%%%%%

\section{\label{sec:five}Heavy-quarkonium spectrum}

Let us now apply the results of the previous sections to the analysis of the
heavy-quarkonium spectrum.
We restrict the analysis to the perturbative corrections, neglecting issues
like nonperturbative contributions in the case of bottomonium and finite-width
effects in the case of the top-antitop system.
This is justified because the problem of large perturbative corrections seems
to be crucial for the heavy-quarkonium theory.
Furthermore, we only consider the zero-orbital-momentum states, with $l=0$,
which are of primary phenomenological interest.

\boldmath
\subsection{Perturbative $\alpha_s^5m_q$ heavy-quarkonium spectrum}
\unboldmath

The ${\cal O}\left(\alpha_s^3\right)$ corrections to the energy levels arise
from several sources:
\begin{itemize}
\item[(i)] matrix elements of the N$^3$LO operators of the effective
Hamiltonian between Cou\-lomb wave functions; 
\item[(ii)] higher iterations of the NLO and NNLO operators of the effective
Hamiltonian in time-independent perturbation theory; 
\item[(iii)] matrix elements of the N$^3$LO instantaneous operators generated
by the emission and absorption of ultrasoft gluons; and
\item[(iv)] retarded ultrasoft contribution.
\end{itemize}
Parts~(i) and (ii) include corrections due to the running of $\alpha_s$ in the
lower-order operators of the effective Hamiltonian proportional to $\beta_i$,
with $i=0,1,2$.
The logarithmic part of these corrections can be taken into account by
choosing the relevant soft normalization scale of $\alpha_s$ in the NNLO
result for the spectrum to be $\mu_s\approx\alpha_sm_q$.
However, there are also nonlogarithmic corrections proportional to the QCD
beta function.
We postpone the calculation of these corrections to a future publication.
Here, we focus our attention on the conceptually interesting non-RG
corrections.
In the absence of the running of $\alpha_s$, the calculation of part~(ii) is
reduced to a redefinition of $\alpha_s$ in the leading Coulomb approximation.
The matrix elements relevant for parts~(i) and (iii) are conveniently
evaluated in coordinate space.
All necessary formulae, including Fourier transforms, can be found in
Ref.~\cite{TitYnd}.
Part~(iv) corresponds to the $\bfm k$-dependent term of Eq.~(\ref{imnres}).
The ultrasoft corrections to the $n$-th energy level are given by
$J_{nn}(E_n)$, and its retarded part can be written as
\begin{equation}
{2C_F^3\alpha_s^3\over3\pi}\left|E^C_n\right|L^E_n,
\end{equation}
where we have introduced the QCD Bethe logarithms \cite{KniPen1}
\begin{equation}
L^E_n={1\over C_F^2\alpha_s^2E^C_n}\int{{\rm d}^3{\bfm k}\over(2\pi)^3}
|\langle{\bfm r}\rangle_{{\bff k}n}|^2
\left(E^C_n-{{\bfm k}^2\over m_q}\right)^3
\ln{E^C_1\over E^C_n-{\bfm k}^2/m_q}.
\end{equation}
The latter can be reduced to one-parameter integrals of elementary functions
\cite{KniPen1}.\footnote{%
There are two misprints in Ref.~\cite{KniPen1}: in Eq.~(A.3), $n$ should be in
the numerator;
in Eq.~(A.5), $\arctan{}$ should be replaced by $\arccot{}$.}
For the reader's convenience, we list the relevant formulae here.
They read
\begin{equation}
L^E_n=\int_0^\infty{\rm d}\nu\,Y_n^E(\nu)X^2_n(\nu),
\end{equation}
where
\begin{eqnarray}
Y^E_n(\nu)&=&{2^6\rho_n^5\nu(\nu^2+1)\exp[4\nu\arccot(\nu/\rho_n)]\over
n^2(\nu^2+\rho_n^2)^3[\exp(2\pi\nu)-1]}\ln{n^2\nu^2\over\nu^2+\rho_n^2},
\nonumber\\
X_1(\nu)&=&\rho_1+2,
\nonumber\\
X_2(\nu)&=&{\nu^2(2\rho_2^2+9\rho_2+8)-\rho_2^2(\rho_2+4)\over
(\nu^2+\rho_2^2)},
\nonumber\\
X_3(\nu)&=&{\nu^4(8\rho_3^3+60\rho_3^2+123\rho_3+66)
-2\nu^2\rho_3^2(6\rho_3^2+41\rho_3+54)+3\rho_3^4(\rho_3+6)\over
3(\nu^2+\rho_3^2)^2},
\end{eqnarray}
with
\begin{equation}
\rho_n=n\left({C_A\over2C_F}-1\right)={n\over8}.
\end{equation}
The expressions for $X_n$ with $n>3$ are usually irrelevant for practical
applications.
For $n=1,2,3$, we obtain the following numerical values:
\begin{equation}
L^E_1=-81.5379,\qquad L^E_2=-37.6710,\qquad L^E_3=-22.4818.
\end{equation}
Putting everything together and writing
\begin{equation}
E_n=E^C_n+\delta E^{(1)}_n+\delta E^{(2)}_n+\delta E^{(3)}_n+\cdots,
\end{equation}
we obtain our final result for the N$^3$LO corrections to the heavy-quarkonium
energy levels in the approximation of putting $\beta(\alpha_s)=0$:
\begin{eqnarray}
\left.\delta E^{(3)}_n\right|_{\beta(\alpha_s)=0}
&=&\left|E^C_n\right|{\alpha_s^3\over\pi}\left\{
-{a_1a_2+a_3\over32\pi^2}
+\left[-{C_AC_F\over2}+\left(-\frac{7}{4}+{9\over16n}
+{S(S+1)\over2}\right)C_F^2\right]{a_1\over n}\right.
\nonumber\\
&&{}+\left[{5\over36}+{1\over6}
\left(\ln2-\gamma_E-\ln n-\Psi_1(n+1)+L_{\alpha_s}\right)\right]C_A^3
\nonumber\\
&&{}+\left[-{97\over36}
+{4\over3}\left(\ln2+\gamma_E-\ln n+\Psi_1(n+1)+L_{\alpha_s}\right)
\right]{C_A^2C_F\over n}
\nonumber\\
&&{}+\left[\left(-{139\over36}+4\ln2+{7\over6}(\gamma_E-\ln n+\Psi_1(n+1))
+{41\over6}L_{\alpha_s}\right)\right.
\nonumber\\
&&{}+\left({47\over24}
+{2\over3}\left(-\ln2+\gamma_E+\ln n+\Psi_1(n+1)-L_{\alpha_s}\right)\right)
{1\over n}
\nonumber\\
&&{}+\left.
\left({107\over108}-{7\over12n}
+{7\over 6}(\gamma_E-\ln{n}+\Psi_1(n+1)-L_{\alpha_s})\right)S(S+1)\right]
{C_AC_F^2\over n}
\nonumber\\
&&{}+\left[{79\over18}-{7\over6n}+{8\over3}\ln2
+{7\over3}(\gamma_E-\ln n+\Psi_1(n+1))+3L_{\alpha_s}-{S(S+1)\over3}\right]
{C_F^3\over n}
\nonumber\\
&&{}+\left[-{32\over15}+2\ln2+(1-\ln2)S(S+1)\right]{C_F^2T_F\over n}
\nonumber\\
&&{}\left.+{49C_AC_FT_Fn_l\over36n}
+\left[{8\over9}-{5\over18n}-{10\over27}S(S+1)\right]{C_F^2T_Fn_l\over n}
+{2\over3}C_F^3L^E_n\right\},
\label{spectrum}
\end{eqnarray}
where $S$ is the spin quantum number,
$L_{\alpha_s}=-\ln(C_F\alpha_s)$,
$\Psi_1(z)=d\Gamma(z)/dz$,
$\Gamma(z)$ is Euler's gamma function,
and $\gamma_E=0.577216\ldots$ is Euler's constant.
The terms proportional to $a_1$ correspond to iterations of lower-order
operators.
We have not included in Eq.~(\ref{spectrum}) the imaginary part corresponding
to the partial width of the decay of the $S=0$ state to two gluons,
\begin{equation}
\Gamma_{gg}={C_F^4T_F\alpha_s^5m_q\over2n^3}.
\end{equation}
The logarithmic part of Eq.~(\ref{spectrum}),
\begin{equation}
\delta E_n^{(3)}=\left|E_n^C\right|{\alpha_s^3\over\pi}
\left\{{1\over6}C_A^3+{4\over3n}C_FC_A^2
+\left[{41\over6n}-{7\over6n}S(S+1)-{2\over3n^2}\right]C_F^2C_A
+{3\over n}C_F^3\right\}\ln{1\over\alpha_s},
\label{log}
\end{equation}
is known from previous analyses \cite{BPSV2,KniPen2}.
For the corrections to the $n=1,2,3$ energy levels, we find numerically
\begin{eqnarray}
\delta E^{(3)}_1&\approx&-{\alpha_s^3\over\pi}\left|E^C_1\right|
\left[{a_3\over32\pi^2}+177.716-11.611\,n_l+0.274\,n_l^2-0.004\,n_l^3\right.
\nonumber\\
&&{}-\left.60.500\,\ln{1\over\alpha_s}+\left(-18.853+1.312\,n_l
+6.222\,\ln{1\over\alpha_s}\right)S(S+1)\right],
  \nonumber
  \\
\delta E^{(3)}_2&\approx&-{\alpha_s^3\over\pi}\left|E^C_2\right|
\left[{a_3\over32\pi^2}+102.917-8.034\,n_l+0.274\,n_l^2-0.004\,n_l^3\right.
\nonumber\\
&&{}-\left.33.389\,\ln{1\over\alpha_s}
+\left(-9.603+0.658\,n_l+3.111\,\ln{1\over\alpha_s}\right)S(S+1)\right],
  \nonumber
  \\
\delta E^{(3)}_3&=&-{\alpha_s^3\over\pi}\left|E^C_3\right|
\left[{a_3\over32\pi^2}+75.919-6.690\,n_l+0.274\,n_l^2-0.004\,n_l^3\right.
\nonumber\\
&&{}-\left.
23.957\,\ln{1\over\alpha_s}
+\left(-6.425+0.439\,n_l+2.074\,\ln{1\over\alpha_s}\right)S(S+1)\right].
\label{num}
\end{eqnarray}

\subsection{Numerical estimates and phenomenological examples}

To illustrate the phenomenological relevance of our results, let us consider
two important physical examples: the resonance in top-antitop threshold
production by $e^+e^-$ annihilation via a virtual photon and the lowest
$\Upsilon$ resonance.
We neglect nonperturbative contributions and finite-width effects, so that the
masses of the resonances are determined by the perturbative expressions with
principal quantum number $n=1$ and spin quantum number $S=1$.
The complete NLO and NNLO corrections may be found in
Refs.~\cite{PinYnd,PenPiv1,MelYel1}, and the N$^3$LO ones for
$\beta(\alpha_s)=0$ are given in Eq.~(\ref{num}).
To take into account the N$^3$LO RG logarithms, we normalize $\alpha_s$ in NLO
and NNLO at the soft scale  $\mu_s=C_F\alpha_s(\mu_s)m_q$.
The setting of the normalization scale in the ${\cal O}(\alpha_s^3)$
corrections is a more subtle problem.
In this order, the hard and ultrasoft regions start to contribute.
This results in RG logarithms with corresponding scales.
Furthermore, the contributions from different regions are not separately
finite, and the operators of the effective Hamiltonian acquire anomalous
dimensions, which result in non-RG logarithms [see Eqs.~(\ref{Coulcor}),
(\ref{nacor}), and (\ref{di})].
Thus, starting with the next order, the RG and non-RG logarithms mix.
The correct treatment of the logarithmic corrections is possible within the
effective-theory RG approach
\cite{BSS2,LMR,ManSte2,PinSot4,HMST,HMS,Pin1,Pin2}.
For simplicity, we ignore these sophistications for the time being and employ
the soft normalization point for the whole of the ${\cal O}(\alpha_s^3)$
corrections.
As an estimate of the nonlogarithmic corrections proportional to $\beta_i$, we
use the $\beta_0^3$ term, which is currently only known for $n=1$
\cite{KiySum},
\begin{equation}
\left.\delta E^{(3)}_1\right|_{\beta_0^3}
=-\left|E^C_n\right|\left({4\alpha_s\beta_0\over\pi}\right)^3
\left[-{1\over8}+{\pi^2\over16}+{\pi^4\over1440}+\zeta(3)
-{\pi^2\over8}\zeta(3)+{3\over2}\zeta(5)\right],
\label{bt0}
\end{equation}
with $\zeta(5)=1.036928\ldots$, and is expected to dominate the corrections of
this type.

For the top-antitop system, we use $m_t=176$~GeV and
$\alpha_s(\mu_s)=0.14$ to find the perturbative expansion of the
binding energy to be
\begin{equation}
E_1=-1.53~\mbox{GeV}\times[1+0.448+0.322
+(0.006+0.011|_{a_3}+0.073|_{\beta_0^3})+\cdots],
\label{top}
\end{equation}
where the N$^3$LO contributions due to the $a_3$ and $\beta_0^3$ terms are
given separately. 
Thus, the resonance peak position is decreased by approximately 140~MeV in
comparison to the NNLO result.
Although the ${\cal O}(\alpha_s^3)$ corrections are still important, the
series shows reasonable
convergence, which makes us optimistic about an accurate
determination of $m_t$ and $\alpha_s$ from this observable.

As for the $\Upsilon(1S)$ resonance, we use $m_b=4.8$~GeV and
$\alpha_s(\mu_s)=0.31$ to find the perturbative expansion of the
binding energy to be
\begin{equation}
E_1=-205~\mbox{MeV}\times[1+1.11+1.88+(0.49+0.19|_{a_3}+1.02|_{\beta_0^3})
+\cdots].
\label{bot}
\end{equation}
This implies that the value of $m_b$ extracted from the $\Upsilon(1S)$
resonance is increased by approximately 170~MeV in comparison to the NNLO
result.
Although there is no further growth of the perturbative corrections, the
${\cal O}(\alpha_s^3)$ corrections seem to be too large to expect a reliable
prediction from the N$^3$LO result, and some optimization, e.g.\ by mass
and/or coupling-constant redefinition, is necessary to improve the convergence
of the series.   

In the above estimates we used the Pad\'e results of Eq.~(\ref{a3}) for the
coefficient $a_3$.
The accuracy of the Pad\'e approximation is difficult to estimate, and a 
significant deviation from the exact result does not seem impossible. 
However, from Eqs.~(\ref{top}) and (\ref{bot}) we observe that the
corresponding contribution only provides about 10\% of the total
${\cal O}(\alpha_s^3)$ corrections, and even a 100\% variation of $a_3$ merely
results in a $10\%$ variation of the ${\cal O}(\alpha_s^3)$ corrections.
This is not crucial for the top-antitop system, but it could be essential in
the bottomonium case, where the magnitude of the ${\cal O}(\alpha_s^3)$
contribution is very sizeable.
The analytical evaluation of the coefficient $a_3$ is thus quite important.

The origin of the large NLO and NNLO corrections in Eqs.~(\ref{top}) and
(\ref{bot}) is usually attributed to the IR-renormalon contribution.
This contribution is absent in the IR-safe ``short-distance'' masses, and the
perturbative series for such masses are expected to exhibit faster convergence
(see, e.g., Refs.~\cite{MelYel2,Hoa,BenSin}). 
We observe, however, that, in the top-antitop case, the perturbative series
numerically converges even in the pole-mass scheme.
On the other hand, in the bottomonium case, the ${\cal O}(\alpha_s^3)$
corrections remain sizeable even for $\beta(\alpha_s)=0$, i.e.\ the na\"\i ve
subtraction of the renormalon contribution through a mass redefinition does
not completely solve the problem of large perturbative corrections. 

Our next comment concerns the corrections logarithmic in $\alpha_s$.
Using the effective-theory RG equations, it is possible to sum up the
logarithmic corrections to the energy levels to all orders.
The presence of several correlated scales renders the problem very interesting
and nontrivial from the conceptual point of view.
Now there are two contradictory results \cite{HMS,Pin1} on the resummation of
the $\alpha_s^{n+4}\ln^n\alpha_s$ terms in the series for the energy levels,
the first of which is given by Eq.~(\ref{log}).
The result of Ref.~\cite{Pin1} is obtained within pNRQCD, while an alternative
formulation of effective theory, velocity NRQCD (vNRQCD), was used in
Ref.~\cite{HMS}.
Since there can be only one correct result, this issue has to be clarified.
In any case, it is interesting to check the accuracy of the logarithmic
approximation.
Numerically, the logarithmic series is dominated by its first term, which
provides about 80\% in both the bottomonium and top-antitop cases
\cite{HMS,Pin1,Pin3}.
From Eq.~(\ref{num}), we observe that the full result has approximately the
same magnitude, but the opposite sign compared to the logarithmic
contribution.
We thus conclude that, while the logarithmic contributions in some order can
give us a hint at the order of magnitude of the full contribution of that 
order, the practical relevance of the high-order resummation is questionable.
This is not unexpected because the resummation parameter
$\alpha_s^n\ln^m\alpha_s$ is neither large for $\alpha_s\approx0$ nor for
$\alpha_s\approx1$.

\section{\label{sec:six}Conclusion}

In this paper, we took a crucial step towards the N$^3$LO analysis of the
heavy-quark threshold dynamics.
We used the effective theory of pNRQCD and the threshold expansion for a
detailed analysis of the nonrelativistic Hamiltonian in this order.
Explicit expressions for the N$^3$LO Hamiltonian in one and two loops
were given.
We also presented the full matching of the Hamiltonian to the contribution
from the ultrasoft gluons, which enter the bound-state dynamics in this order.
The matching calculation includes one-, two-, and three-loop operators.
To complete our analysis, the three-loop $\overline{\rm MS}$ coefficient $a_3$
of the corrections to the static potential, for which only Pad\'e estimates 
are available, has to be computed.

With the full expression for the Hamiltonian and the ultrasoft contribution at
hand, it is straightforward to complete the N$^3$LO analysis of the
heavy-quarkonium spectrum.
In this paper, we derived the heavy-quarkonium spectrum in this order,
neglecting the nonlogarithmic terms proportional to the QCD beta function.
For the latter, only the $\beta_0^3$ term for the ground-state energy is known
so far.

Collecting all available contributions, we found the N$^3$LO corrections to be
sizable for the top-antitop system, where, however, the perturbative series
for the resonance-peak energy exhibits a tendency to converge even in the
pole-mass scheme.
In the case of the $\Upsilon(1S)$ resonance, the corrections are so sizeable,
that some kind of optimization of the perturbative expansion is needed, e.g.\
by removing the pole mass in favour of a renormalon-free short-distance mass.
However, we should emphasize that, in the bottomonium case, the N$^3$LO
corrections remain sizeable even for $\beta(\alpha_s)=0$, and the bad
behaviour of the perturbative series cannot be solely explained by the
renormalon contribution.

In order to render the analysis more accurate, the remaining nonlogarithmic
terms proportional to the QCD beta function have to be evaluated along with
the $a_3$ coefficient. However, we do not expect a qualitative 
change of our result.

The result of this paper also provides a starting point for the calculation of
the N$^3$LO single-logarithmic $\alpha_s^3\ln\alpha_s$ corrections to the
heavy-quarkonium production and annihilation rates, similarly to a number of
QED bound-state problems \cite{KniPen3,HilLep,MelYel3,Hil}.
While the analysis of the nonlogarithmic terms in this order requires the
calculation of the three-loop hard renormalization of the relevant production
and annihilation amplitudes, which still is a challenging theoretical problem,
the logarithmic terms are universal and essentially determined by the
effective Hamiltonian and the ultrasoft contribution presented in this paper.
Along with the known double-logarithmic $\alpha_s^3\ln^2\alpha_s$ terms
\cite{KniPen2}, the single-logarithmic contribution would constitute an
essential part of the N$^3$LO corrections to the heavy-quarkonium production
and annihilation rates.
The calculation of the single-logarithmic terms would also lead to further
progress in the resummation of the logarithms in $v$ via
the nonrelativistic effective-theory RG \cite{HMST,Pin2}, since it determines
an anomalous dimension necessary for the NNLO logarithmic analysis of
heavy-quarkonium production and annihilation.
Although it is, in general, dangerous to rely on the logarithmic
approximation, as we observed in the case of the spectrum, the situation could
be different for the cross section normalization, where the N$^3$LO
double-logarithmic contribution is known to be sizeable and the resummation of
the logarithmic terms could stabilize the behavior of perturbation theory
\cite{HMST}.

\bigskip
\noindent
{\bf Acknowledgements}
\smallskip

\noindent
A.A.P. is grateful to A.H. Hoang and A. Pineda for useful communications.
We thank J. Soto for helpful comments on this manuscript.
This work was supported in part by the Deutsche Forschungsgemeinschaft through
Grant No.\ KN~365/1-1 and by the Bundesministerium f\"ur Bildung und Forschung
through Grant No.\ 05 HT1GUA/4.
The work of V.A.S. was supported in part by the Russian Foundation for Basic
Research through Project No.\ 01-02-16171 and by INTAS 
through Grant No.\ 00-00313.

\section*{Appendix}

To evaluate the two-loop contribution to the heavy-quark potential, in
particular the $1/m_q$ corrections, one needs analytical results, at least
as Laurent expansions in $\epsilon$ up to some order (typically,
$\epsilon^0$ and $\epsilon^1$), for the following family of two-loop Feynman
integrals:
\begin{eqnarray}
J(a_1,\ldots,a_8;q^2;\epsilon)&=&
\int\int\frac{{\rm d}^dk{\rm d}^dl}{(k^2)^{a_1}(l^2)^{a_2}
[(k-q)^2]^{a_3}[(l-q)^2]^{a_4}[(k-l)^2]^{a_5}}
\nonumber\\
&&{}\times\frac{1}{(v_0\cdot k)^{a_6}(v_0\cdot l)^{a_7}[v_0\cdot(k-l)]^{a_8}},
\label{Jint}
\end{eqnarray}
where the four-vector $v_0$ is defined below Eq.~(\ref{prop}), $k$ and $l$ are
loop four-momenta, and $+i\varepsilon$ is omitted in all the denominators. 

As in Ref.~\cite{Pet,Sch,Sch1}, we use a reduction procedure that expresses
any integral of the form of Eq.~(\ref{Jint}) in terms of some master 
integrals.
To this end, we employ the following identities, which can be obtained by 
means of the method of integration by parts \cite{CheTka},
\begin{eqnarray}
(d-2a_1-a_3-a_5-a_6)
+a_3\bfm{3^+}(q^2-\bfm{1^-})
-a_5\bfm{5^+}(\bfm{1^-}-\bfm{2^-})
-a_8\bfm{6^-8^+}&=&0,
\nonumber\\
(d-2a_2-a_3-a_5-a_7)
+a_4\bfm{4^+}(q^2-\bfm{2^-})
-a_5\bfm{5^+}(\bfm{2^-}-\bfm{1^-})
+a_8\bfm{7^-8^+}&=&0,
\nonumber\\
(d-a_1-a_3-2a_5-a_6-a_8)
+a_1\bfm{1^+}(\bfm{2^-}-\bfm{5^-})
+a_3\bfm{3^+}(\bfm{4^-}-\bfm{5^-})
+a_6\bfm{6^+7^-}&=&0,
\nonumber\\
(d-a_2-a_4-2a_5-a_7-a_8)
+a_2\bfm{2^+}(\bfm{1^-}-\bfm{5^-})
+a_4\bfm{4^+}(\bfm{3^-}-\bfm{5^-})
+a_7\bfm{6^-7^+}&=&0,
\nonumber\\
(d-a_1-2a_3-a_5-a_6)
+a_1\bfm{1^+}(q^2-\bfm{3^-})
-a_5\bfm{5^+}(\bfm{3^-}-\bfm{4^-})
-a_8\bfm{6^-8^+}&=&0,
\nonumber\\
(d-a_2-2a_4-a_5-a_7)
+a_2\bfm{2^+}(q^2-\bfm{4-})
-a_5\bfm{5^+}(\bfm{4^-}-\bfm{3^-})
+a_8\bfm{7^-8^+}&=&0,
\nonumber\\
2a_1\bfm{1^+6^-}
+2a_3\bfm{3^+6^-}
+a_5\bfm{5^+8^-}
+v^2a_6\bfm{6^+}
+v^2a_8\bfm{8^+}&=&0,
\nonumber\\
2a_2\bfm{2^+7^-}
+2a_4\bfm{4^+7^-}
-a_5\bfm{5^+8^-}
+v^2a_7\bfm{7^+}
-v^2a_8\bfm{8^+}&=&0,\quad
\end{eqnarray}
as well as the trivial identity $\bfm{6^-}-\bfm{7^-}=\bfm{8^-}$.
Here, the standard notation for raising and lowering operators has been used,
e.g.
\begin{equation}
\bfm{1^-3^+}J(a_1,\ldots,a_8)=J(a_1-1,a_2,a_3+1,\ldots,a_8).
\end{equation}
We developed a reduction procedure very similar to the one of
Ref.~\cite{Sch1}.
In our problem, however, we need a larger class of integrals that arise in
the calculation of the $1/m_q$ corrections in the general covariant gauge.
The main difference between our reduction procedure and the one of
Ref.~\cite{Sch1} is that we stop the reduction if we arrive at integrals
expressed in terms of gamma functions for finite $\epsilon$.
There are two subclasses of the integrals of Eq.~(\ref{Jint}) that are only
evaluated as expansions in $\epsilon$ up to some order.
The first of them was described in Ref.~\cite{Pet}, namely
\begin{equation}
I(a_1,\ldots,a_5;q^2;\epsilon)
=\int\int\frac{{\rm d}^dk{\rm d}^dl}{(k^2)^{a_1}(l^2)^{a_2}
[(k-l-q)^2]^{a_3}(v_0\cdot k)^{a_4}(v_0\cdot l)^{a_5}}.
\label{Iint}
\end{equation}
In particular, we have
\begin{eqnarray}
J(0, a_2, a_3, 0, a_5, a_6, 0, a_8) &=& I(a_5, a_3, a_2, a_8, a_6),
\nonumber\\
J(a_1, 0, 0, a_4, a_5, a_6, a_7, 0)& = &I(a_1, a_4, a_5, a_6, a_7).
\end{eqnarray}
The master integrals for this subclass are
\begin{eqnarray}
I(1,1,1,1,1;q^2;\epsilon)&=&
\frac{\left(i\pi^{d/2}{\rm e}^{-\gamma_E\epsilon}\right)^2}
{(-q^2)^{2\epsilon}}
\left[-\frac{2\pi^2}{3\epsilon}-4\pi^2
-\left(24\pi^2-\frac{7\pi^4}{9}\right)\epsilon+{\cal O}(\epsilon^2)\right],
\nonumber\\
I(1,1,2,1,1;q^2;\epsilon)&=&
\frac{\left(i\pi^{d/2}{\rm e}^{-\gamma_E\epsilon}\right)^2}
{(-q^2)^{1+2\epsilon}}
\left[\frac{2}{\epsilon^2}-\frac{4}{\epsilon}+8-\frac{5\pi^2}{3}
-\left(16-\frac{10\pi^2}{3}+\frac{64\zeta(3)}{3}\right)\epsilon\right.
\nonumber\\
&&{}+\left.{\cal O}(\epsilon^2)\right].
\end{eqnarray}
We need also integrals of the type of Eq.~(\ref{Iint}) with a numerator that
can be chosen to be $q\cdot k$ or $q\cdot l$.
The reduction of these integrals is quite similar and also results in two
master integrals.

The second subclass of the integrals of Eq.~(\ref{Jint}) that are not
expressed in terms of gamma functions for general $\epsilon$ consists of
integrals with $a_5=a_6=a_7=0$.
Using Feynman parameters, the general integral of this kind can be represented
in terms of the following Mellin-Barnes representation:
\begin{eqnarray}
\lefteqn{J(a_1,\ldots,a_4,0,0,0,a_5)=
\frac{(-1)^a2^{a_5-1}\left(i\pi^{d/2}\right)^2}
{\prod_{i=1}^5\Gamma(a_i)(-q^2)^{a-a_5/2-4+2\epsilon}}}
\nonumber\\
&&{}\times\frac{1}{2\pi i} \int_{-i\infty}^{+i\infty}{\rm d}z
\frac{\Gamma(a_1 + a_2 + a_3 + a_4 + a_5/2 - 4 + 2 \epsilon + z)
\Gamma(a_1 + z) \Gamma(a_3 + z)}{\Gamma(a_1 + a_3 + 2z)\Gamma(-z)}
\nonumber\\
&&{}\times\frac{\Gamma(-a_1 - a_3 - a_4 - a_5/2 + 4 - 2\epsilon - z)
\Gamma(-a_1 - a_2 - a_3 - a_5/2 + 4 - 2\epsilon - z)}
{\Gamma(-2a_1 - a_2 - 2a_3 - a_4 - a_5 + 8 - 4\epsilon - 2z)}
\nonumber\\
&&{}\times
\Gamma(a_1 + a_3 + a_5/2 - 2 + \epsilon + z)
\Gamma(-a_1 - a_3 + 2 - \epsilon - z),
\label{YSMB}
\end{eqnarray}
where $a=\sum_{i=1}^5a_i$.
We performed a reduction to integrals with $a_2=a_3=a_4=a_5=1$ and evaluated
them by means of Eq.~(\ref{YSMB}) for the required values of $a_1$.
In particular, we found
\begin{equation}
J(1,1,1,1,0,0,0,1;q^2;\epsilon)
=\frac{\left(i\pi^{d/2}{\rm e}^{-\gamma_E\epsilon}\right)^2}
{(-q^2)^{1/2+2\epsilon}}
\left\{-4\pi^2\ln2[1+\epsilon(4+\ln2)]+\frac{5\pi^4}{3}\epsilon
+{\cal O}(\epsilon^2)\right\}.
\end{equation}

\end{document}